\documentclass[12pt]{article}
\usepackage{mathrsfs}
\usepackage{graphicx}
\usepackage{amsmath}
\usepackage{amssymb}
\usepackage{caption2}
\begin{document}
\newcommand  {\ba} {\begin{eqnarray}}
\newcommand  {\be} {\begin{equation}}
\newcommand  {\ea} {\end{eqnarray}}
\newcommand  {\ee} {\end{equation}}
\renewcommand{\thefootnote}{\fnsymbol{footnote}}
\renewcommand{\figurename}{Figure.}
\renewcommand{\captionlabeldelim}{.~}

\vspace*{1cm}
\begin{center}
 {\Large\textbf{A Model of Asymmetric Hadronic Dark Matter and Leptogenesis}}

\vspace{1cm}
 \textbf{Wei-Min Yang}

\vspace{0.3cm}
 \emph{Department of Modern Physics, University of Science and Technology of China, Hefei 230026, P. R. China}

\vspace{0.3cm}
 \emph{E-mail: wmyang@ustc.edu.cn}
\end{center}

\vspace{1cm}
\noindent\textbf{Abstract}: The paper suggests a model to account for the common origins of the asymmetric dark matter (ADM) and matter-antimatter asymmetry. The ADM nature is a stable hadronic particle consisting of a heavy color scalar and a light $u$ quark, which is formed after the QCD phase transition. At the early stage the ADM are in thermal equilibrium through collisions with the nucleons, moreover, they can emit the $\gamma$ photons with $0.32$ MeV energy. However they are decoupling and become the dark matter at the temperature about $130$ MeV. The mass upper limit of the ADM is predicted as $M_{D}<1207$ GeV. It is feasible and promising to test the model in future experiments.

\vspace{1cm}

\vspace{0.3cm}

\newpage
 \noindent\textbf{I. Introduction}
\vspace{0.3cm}

  The current universe observations and the standard model (SM) of particle physics have established the data of the baryon asymmetry and the dark matter (DM) abundance as follows \cite{1},
\ba
 \eta_{B}=\frac{n_{B}-\overline{n}_{B}}{n_{\gamma}}\approx6.1\times10^{-10},\hspace{0.5cm}
 \frac{\Omega_{D}}{\Omega_{B}}\approx 5\,.
\ea
How did the two values originate in the universe evolution? Is there a relationship between them? What is truly the DM nature? What clues are left to us for the DM detections? All the time the issues attract great attentions in the fields of experiment and theory because they are very significant for particle physics and cosmology \cite{2}. Although the SM is a very successful theory at the electroweak energy scale \cite{3}, it can not account for the cosmic problems. All of these are unsolved puzzles up to now \cite{4}. Undoubtedly, we need the new physics beyond the SM in order to understand the matter origin and the early universe evolution well.

  All kinds of theoretical ideas have been suggested to solve the above-mentioned problems. The baryon asymmetry can be achieved by the electroweak baryogenesis \cite{5}, the thermal leptogenesis \cite{6}, and so on. The candidates of the cold dark matter are possibly the real scalar boson \cite{7}, the sterile neutrino \cite{8}, the lightest supersymmetric particle \cite{9}, the axion \cite{10}, and so on. Recently, the asymmetric dark matter is a well-motivated idea because it has something to do with the baryon asymmetry \cite{11}. The common origin of matter and dark matter was studied in the reference \cite{12}. On the basis of the unified whole of nature, a realistic theory beyond the SM should simultaneously accommodate and account for the neutrino physics, baryon asymmetry and dark matter besides the SM, in other words, it has to integrate the four things completely. It is especially hard for a model construction to keep the principle of simplicity, feasibility and fewer number of parameters, otherwise, the theory will be excessive complexity so that it is incredible or infeasibility. Although a great deal of efforts have been made toward the ultimate solutions, it is still a large challenge for theoretical particle physicists to realize the purpose \cite{13}.

  In this work, I suggest a simple and feasible particle model. It is based on the SM gauge groups but appends a $Z_{2}$ discrete symmetry. Besides the SM particles, two super-heavy Majorana fermions and three color scalar bosons are new introduced in the model. The baryon and DM asymmetries stem from the decay chains of the two Majorana fermions in common. The DM nature is an asymmetric hadronic particle which consists of the lightest color scalar and a $u$ quark after the QCD phase transition. They can give out light in the stage of the thermal equilibrium through collisions with the nucleons. After they are decoupling, they really become the dark matter. The model predicts three interesting results of the DM, namely the transition energy of $0.32$ MeV, the decoupling temperature of $130$ MeV, and the mass upper limit of $1207$ GeV. Finally, the model is feasible and promising to be tested in future experiments.

  The remainder of this paper is organized as follows. In Section II I outline the model. Sec. III I discuss the asymmetric hadronic dark matter. Sec. IV I give the numerical results and the experimental searches. Sec. V is devoted to conclusions.

\vspace{1cm}
 \noindent\textbf{II. Model}
\vspace{0.3cm}

  The model symmetries are characterized together by the SM gauge groups, the global baryon number conservation $U(1)_{B}$, and a discrete symmetry $Z_{2}$, namely $SU(3)_{C}\otimes SU(2)_{L}\otimes U(1)_{Y}\otimes U(1)_{B}\otimes Z_{2}$. Under $Z_{2}$, all of the gauge bosons and right-handed fermions have ``$+1$" parity, and all of the scalar bosons and left-handed fermions have ``$-1$" parity. The model particle contents include the whole SM particles and the following new particles,
\ba
 N_{R}(1,1,0),\hspace{0.3cm} \chi_{L}(1,1,0),\hspace{0.3cm} \Phi_{L}(3,2,\frac{1}{3}),\hspace{0.3cm} \Phi_{R1}(3,1,-\frac{2}{3}),\hspace{0.3cm} \Phi_{R2}(3,1,\frac{4}{3}).
\ea
These numbers in the parentheses are the gauge quantum numbers. $N_{R}$ and $\chi_{L}$ are two gauge singlets but they have opposite parities under $Z_{2}$, however, they are all Majorana fermions. $\Phi_{L},\Phi_{R1},\Phi_{R2}$ are all scalar bosons with the color quantum numbers, so all of them have the baryon number $\frac{1}{3}$. $\Phi_{L}$ is an isospin doublet, while $\Phi_{R1}$ and $\Phi_{R2}$ are isospin singlets. All of the new particles play key roles in the new physics beyond the SM, in particular, in the early universe evolution.

  On the basic of the model symmetries, the full Lagrangian is
\begin{alignat}{1}
 \mathscr{L}=
&\:\mathscr{L}_{SM}+i\overline{N_{R}}\gamma^{\mu}\partial_{\mu}N_{R}
 +i\overline{\chi_{L}}\gamma^{\mu}\partial_{\mu}\chi_{L}\nonumber\\
&+(D^{\mu}\Phi_{L})^{\dagger}(D_{\mu}\Phi_{L})+(D^{\mu}\Phi_{R1})^{\dagger}(D_{\mu}\Phi_{R1})
 +(D^{\mu}\Phi_{R2})^{\dagger}(D_{\mu}\Phi_{R2})\nonumber\\
&-(\frac{1}{2}N_{R}^{T}CM_{N}N_{R}+\frac{1}{2}\chi_{L}^{T}CM_{\chi}\chi_{L}+h.c.)\nonumber\\
&-(\overline{l}\widetilde{H}Y_{N}N_{R}+\overline{q}\Phi_{L}Y_{L}N_{R}
 +\overline{d_{R}}\Phi_{R1}Y_{1}\chi_{L}+\overline{u_{R}}\Phi_{R2}Y_{2}\chi_{L}+h.c.)\nonumber\\
&-(M_{L}^{2}\Phi_{L}^{\dagger}\Phi_{L}+M_{1}^{2}\Phi_{R1}^{\dagger}\Phi_{R1}
 +M_{2}^{2}\Phi_{R2}^{\dagger}\Phi_{R2})\nonumber\\
&-(\lambda_{L}\Phi_{L}^{\dagger}\Phi_{L}+\lambda_{1}\Phi_{R1}^{\dagger}\Phi_{R1}
 +\lambda_{2}\Phi_{R2}^{\dagger}\Phi_{R2})H^{\dagger}H
 -(\lambda_{12}\Phi_{R1}^{\dagger}\Phi_{R2}H^{\dagger}\widetilde{H}+h.c.)\nonumber\\
&-\mbox{the quartic couplings of } \Phi_{L},\Phi_{R1},\Phi_{R2},
\end{alignat}
where $C$ is a charge conjugation matrix and the self-explanatory notations, $l,q,d_{R}$, $u_{R},H,\widetilde{H}=i\tau_{2}H^{*}$, are the SM particle states. Obviously, the baryon number conservation is incidental in (3), whereas the lepton number is not conserved. All the mass terms of the new particles are directly permitted by the model symmetries. The new particle masses and the scalar coupling parameters are assumed to be in the areas as follows,
\begin{alignat}{1}
&\langle H\rangle=174\:\mbox{GeV}<M_{2}\lesssim M_{1}\sim 10^{3}\:\mbox{GeV}
 \ll M_{L}\sim10^{8}\:\mbox{GeV}\nonumber\\
&\ll M_{N}\sim M_{\chi}\sim 10^{12}\:\mbox{GeV}<T_{reheat}\sim 10^{13}\:\mbox{GeV},\nonumber\\
&0<(\lambda_{L},\lambda_{1},\lambda_{2},\lambda_{12})\sim 0.1,
\end{alignat}
where $T_{reheat}$ is the reheating temperature after the universe inflation. Obviously, the scalar potential in (3) keeps the electroweak vacuum stability. After the electroweak breaking, the light effective neutrino mass is given through the see-saw mechanism as \cite{14a}
\ba
 m_{\nu}=-Y_{N}\frac{\langle H\rangle^{2}}{M_{N}}Y_{N}^{T}.
\ea
The current experiments have established $m_{\nu}\sim 0.01$ eV \cite{14b}, so this implies the constraint relation $Y_{N}M_{N}^{-1}Y_{N}^{T}\sim 10^{-15}$ GeV$^{-1}$. However, the other Yukawa couplings, $Y_{L},Y_{1},Y_{2}$, have large freedoms and are undetermined. Finally, it should be noted that the coupling term $\Phi_{R1}^{\dagger}\Phi_{R2}H^{\dagger}\widetilde{H}$ will automatically vanish after the electroweak breaking.

  After the universe inflation, the reheating temperature can reach to $T_{reheat}\sim 10^{13}$ GeV for most of the inflation models \cite{15}. Therefore there are an immense amount of the super-heavy Majorana fermions $N_{R}$ and $\chi_{L}$ in the reheated universe. Their decays have significant impact on the universe evolution. In the light of (3), there are the decay chains as follows,
\begin{alignat}{1}
&N_{R}\rightarrow l+\widetilde{H}^{*},\hspace{0.5cm} N_{R}\rightarrow q+\Phi_{L}^{*},\hspace{0.5cm}
 \chi_{L}\rightarrow d_{R}+\Phi_{R1}^{*},\hspace{0.5cm} \chi_{L}\rightarrow u_{R}+\Phi_{R2}^{*},\nonumber\\
&\Phi_{L}^{*}\rightarrow \overline{q}+\overline{l}+\widetilde{H},\hspace{0.5cm}
 \Phi_{R1}^{*}\rightarrow \overline{d_{R}}+\overline{u_{R}}+\Phi_{R2},\hspace{0.5cm}
 \Phi_{R1}^{*}\rightarrow \Phi_{R2}^{*}+H+\widetilde{H}^{*}.
\end{alignat}
$N_{R}$ and $\chi_{L}$ are decoupling as the universe temperature falls below $M_{N}$ and $M_{\chi}$. The produced scalar bosons, $\Phi_{L},\Phi_{R1},\Phi_{R2}$, have different fates afterwards. The heavier $\Phi_{L}^{*}$ has only a decay channel, namely it slowly decays into the SM particles via the effective coupling $q^{T}C\Phi_{L}^{*}Y_{L}^{*}M_{N}^{*-1}Y_{N}^{\dagger}\widetilde{H}^{\dagger}l$. $\Phi_{R1}^{*}$ has two decay modes. It can slowly decay into $\Phi_{R2}$ through the effective coupling $d_{R}^{T}C\Phi_{R1}^{*}Y_{1}^{*}M_{\chi}^{*-1}Y_{2}^{\dagger}\Phi_{R2}^{\dagger}u_{R}$, or it can rapidly decay into $\Phi_{R2}^{*}$ via the scalar coupling. Obviously, the former decay rate is far smaller than the later one, so we can completely ignore the former decay mode. The lightest $\Phi_{R2}^{*}$ can not decay at all, however, it is a stable particle.

  The decay processes of $N_{R}$ and $\chi_{L}$ have the following features. Firstly, the irremovable complex phases in the coupling matrices, $Y_{N},Y_{L},Y_{1},Y_{2}$, are explicitly sources of the $CP$ violation. They can lead to a $CP$ asymmetry in each decay process through the interference between the tree diagram and the one-loop ones \cite{16}. The $CP$ asymmetries are defined and calculated as follow,
\begin{alignat}{1}
&\varepsilon_{1}=\frac{\Gamma_{1}(N_{i}\rightarrow l+\widetilde{H}^{*})
 -\overline{\Gamma}_{1}(N_{i}\rightarrow \overline{l}+\widetilde{H})}
 {\Gamma_{1}+\overline{\Gamma}_{1}+\Gamma_{2}+\overline{\Gamma}_{2}}
 =\frac{\sum\limits_{j\neq i}f(x_{j})\mbox{Im}(Y_{N}^{\dagger}Y_{N})_{ij}^{2}}{8\pi[(Y_{N}^{\dagger}
 Y_{N})_{ii}+3(Y_{L}^{\dagger}Y_{L})_{ii}]}\,,\nonumber\\
&\varepsilon_{2}=\frac{\Gamma_{2}(N_{i}\rightarrow q+\Phi_{L}^{*})
 -\overline{\Gamma}_{2}(N_{i}\rightarrow \overline{q}+\Phi_{L})}
 {\Gamma_{1}+\overline{\Gamma}_{1}+\Gamma_{2}+\overline{\Gamma}_{2}}
 =\frac{3\sum\limits_{j\neq i}f(x_{j})\mbox{Im}(Y_{L}^{\dagger}Y_{L})_{ij}^{2}}{8\pi[(Y_{N}^{\dagger}
 Y_{N})_{ii}+3(Y_{L}^{\dagger}Y_{L})_{ii}]}\,,\nonumber\\
&\varepsilon_{3}=\frac{\Gamma_{3}(\chi_{i}\rightarrow d+\Phi_{R1}^{*})
 -\overline{\Gamma}_{3}(\chi_{i}\rightarrow \overline{d}+\Phi_{R1})}
 {\Gamma_{3}+\overline{\Gamma}_{3}+\Gamma_{4}+\overline{\Gamma}_{4}}
 =\frac{\sum\limits_{j\neq i}f(y_{j})\mbox{Im}(Y_{1}^{\dagger}Y_{1})_{ij}^{2}}{8\pi[(Y_{1}^{\dagger}
 Y_{1})_{ii}+(Y_{2}^{\dagger}Y_{2})_{ii}]}\,,\nonumber\\
&\varepsilon_{4}=\frac{\Gamma_{4}(\chi_{i}\rightarrow u+\Phi_{R2}^{*})
 -\overline{\Gamma}_{4}(\chi_{i}\rightarrow \overline{u}+\Phi_{R2})}
 {\Gamma_{3}+\overline{\Gamma}_{3}+\Gamma_{4}+\overline{\Gamma}_{4}}
 =\frac{\sum\limits_{j\neq i}f(y_{j})\mbox{Im}(Y_{2}^{\dagger}Y_{2})_{ij}^{2}}{8\pi[(Y_{1}^{\dagger}
 Y_{1})_{ii}+(Y_{2}^{\dagger}Y_{2})_{ii}]}\,,\nonumber\\
&f(x_{j})=\sqrt{x_{j}}[1-(1+x_{j})ln\frac{1+x_{j}}{x_{j}}+\frac{1}{1-x_{j}}],\hspace{0.3cm} x_{j}=\frac{M_{N_{j}}^{2}}{M_{N_{i}}^{2}}\,,\hspace{0.3cm} y_{j}=\frac{M_{\chi_{j}}^{2}}{M_{\chi_{i}}^{2}}\,.
\end{alignat}
Secondly, the decay processes are out-of-equilibrium if the decay rates are much smaller than the Hubble expansion rate of the universe, namely
\begin{alignat}{1}
&\Gamma_{1}=\frac{M_{N_{i}}}{16\pi}(Y_{N}^{\dagger}Y_{N})_{ii}\ll H(T=M_{N_{i}}),\hspace{0.3cm}
 \Gamma_{2}=\frac{3M_{N_{i}}}{16\pi}(Y_{L}^{\dagger}Y_{L})_{ii}\ll H(T=M_{N_{i}}),\nonumber\\
&\Gamma_{3}=\frac{3M_{\chi_{i}}}{32\pi}(Y_{1}^{\dagger}Y_{1})_{ii}\ll H(T=M_{\chi_{i}}),\hspace{0.3cm}
 \Gamma_{4}=\frac{3M_{\chi_{i}}}{32\pi}(Y_{2}^{\dagger}Y_{2})_{ii}\ll H(T=M_{\chi_{i}}),\nonumber\\
 &H(T)=\frac{1.66\sqrt{g_{*}}\,T^{2}}{M_{pl}}\,,
\end{alignat}
where $M_{pl}=1.22\times10^{19}$ GeV and $g_{*}$ is an effective number of relativistic degrees of freedom at the temperature $T$. At $T\approx M_{N_{i}}\approx M_{\chi_{i}}$, the non-relativistic particles are only $N_{i}$ and $\chi_{i}$ in the model, the rest of the model particles are all relativistic states, so one can easily figure out $g_{*}=130.75$ in (8). It is not actually difficult to satisfy (8) as long as the Yukawa couplings are sufficient sizes. Lastly, the decay processes conserve the baryon numbers but violate the lepton number. In short, the decays of $N_{R}$ and $\chi_{L}$ satisfy two items of Sakharov's three conditions \cite{17}, namely $CP$ violation and out-of-equilibrium. As a result, asymmetric number densities of the final state particles and their antiparticles are generated, in addition, the net lepton number is non-vanishing although the net baryon number is still nought.

  As the universe expansion and cooling, $\Phi_{L},\Phi_{R1},\Phi_{R2}$ will become non-relativistic particles in sequence. By virtue of the generated $\Phi_{L}^{*}$ asymmetry, $\Phi_{L}^{*}\rightarrow \overline{q}+\overline{l}+\widetilde{H}$ will contribute a lepton asymmetry which is opposite to one of $N_{R}\rightarrow l+\widetilde{H}^{*}$, thus the net lepton asymmetry is actually a result of both cancellation. When the universe expansion rate falls to being equal to the decay rate of $\Phi_{L}^{*}$, the $\Phi_{L}^{*}$ decay is completed, in other words, it's lifetime has expired. After this the lepton asymmetry has no further variation and is frozen out. Therefore, the freezing temperature $T_{LF}$ is determined by the relation
\begin{alignat}{1}
 \Gamma_{\Phi_{L}}=\frac{M_{L}^{3}}{768\pi^{3}}
 \mbox{Tr}M_{N}^{*-1}Y_{N}^{\dagger}Y_{N}M_{N}^{-1}Y_{L}^{T}Y_{L}^{*}=H(T_{LF}).
\end{alignat}
However $T_{LF}>\langle H\rangle$ is required by the model self-consistency. The later numerical calculations show $T_{LF}\sim 1$ TeV. At this temperature, the relativistic particles are exactly ones of the SM, so $g_{*}(T_{LF})=106.75$ in (9).

  Thing happened next is that $\Phi_{R1}^{*}$ rapidly decays into $\Phi_{R2}^{*}$, accordingly, the $\Phi_{R1}^{*}$ asymmetry is completely transferred into the total asymmetry of $\Phi_{R2}^{*}$. $\Phi_{R2}^{*}$ is at the end of the decay chains, however, it is a stable particle. On account of the baryon number conservation, the total asymmetries of the up-type and down-type quarks in (6) is opposite to one of $\Phi_{R2}^{*}$.

  In the temperature region of $\langle H\rangle<T<T_{LF}$, the sphaleron processes are smoothly put into effect \cite{18}, by which the lepton asymmetry is eventually converted into the baryon asymmetry. At $T=\langle H\rangle$, the electroweak breaking occurs, and then the SM particle masses are generated. At the QCD phase transition temperature $T_{QCD}\approx 220$ MeV, the second and third generation heavy quarks have decayed into the first generation light quarks via the weak interaction and flavour mixing. A part of the asymmetric $u$ and $d$ in the plasma thermal bath are captured by the asymmetric $\Phi_{R2}^{*}$ via the strong interaction to form the asymmetric hadronic particles as follows,
\ba
 u+\Phi_{R2}^{*}\rightarrow \Phi_{u}^{0}=\mbox{DM},\hspace{0.5cm} d+\Phi_{R2}^{*}\rightarrow \Phi_{d}^{-}\rightarrow \Phi_{u}^{0}+e^{-}+\overline{\nu_{e}}\,,
\ea
while the rest of the asymmetric $u$ and $d$ are combined into the asymmetric nucleons, i.e. protons and neutrons. Obviously, $\Phi_{u}^{0}$ and $\Phi_{d}^{-}$ are unconventional hadrons. They are fermions with spin $\frac{1}{2}$ and isospin $\frac{1}{2}$. $\Phi_{u}^{0}$ is a stable particle as a proton, while $\Phi_{d}^{-}$ is similar to a neutron, which can decay into $\Phi_{u}^{0}$ via the weak interaction. $\Phi_{u}^{0}$ is a neutral charge and color singlet, and has vanishing baryon number, in particular, it only takes part in the weak interaction. In a word, the asymmetric $\Phi_{u}^{0}$ is namely the cold dark matter in the model. Because the $\Phi_{R2}^{*}$ mass is far larger than the $u$ mass and the contained gluon potential, the $\Phi_{u}^{0}$ mass is very close to the $\Phi_{R2}^{*}$ one, namely $M_{D}\approx M_{2}$.

  It is well known that the symmetric parts of matter and antimatter eventually annihilate into photons in the universe evolution, while the asymmetric parts are surviving up to now. The above discussions are collected together, then the asymmetries of baryon and DM are given by the relations as follow,
\begin{alignat}{1}
&Y_{B}=\frac{n_{B}-\overline{n}_{B}}{s}=0,\hspace{0.5cm} Y_{L}=2\frac{n_{l}-\overline{n}_{l}}{s}
 =2(\kappa_{1}\frac{\varepsilon_{1}}{g_{*}}-\kappa_{2}\frac{\varepsilon_{2}}{g_{*}}),\nonumber\\
&Y_{D}=\frac{n_{\Phi_{R2}^{*}}-\overline{n}_{\Phi_{R2}^{*}}}{s}
 =\kappa_{3}\frac{\varepsilon_{3}}{g_{*}}+\kappa_{4}\frac{\varepsilon_{4}}{g_{*}}\,,\nonumber\\
&\eta_{B}=7.04c_{s}(Y_{B}-Y_{L}),\hspace{0.5cm}
 \eta_{D}=\frac{n_{D}-\overline{n}_{D}}{n_{\gamma}}=7.04Y_{D},\hspace{0.5cm}
 \frac{\Omega_{D}}{\Omega_{B}}=\frac{M_{D}\eta_{D}}{m_{n}\eta_{B}}\,,
\end{alignat}
where $g_{*}=130.75$. $\kappa_{1,2,3,4}$ are four dilution factors corresponding to the four decay processes, which are related to departure degree from thermal equilibrium. 7.04 is a ratio of the entropy density $s$ to the photon number density $n_{\gamma}$. $c_{s}=\frac{28}{79}$ is a coefficient of the sphaleron conversion. $m_{n}$ is a nucleon mass. In conclusion, the model clearly shows origins of the baryon asymmetry and the asymmetric dark matter, and the close relationship of the both.

\vspace{1cm}
 \noindent\textbf{III. Asymmetric Hadronic Dark Matter}
\vspace{0.3cm}

  After the QCD phase transition and before the BBN beginning, namely in the period of $1\:\mbox{MeV}\lesssim T\lesssim 220\:\mbox{MeV}$, the universe particles include the non-relativistic $\Phi_{u}^{0}$ and nucleons, and the relativistic electrons, neutrinos and photons. In view of $\Phi_{u}^{0}$ appearing, this period is in fact divided into two stages. In the first stage, $\Phi_{u}^{0}$ are in thermal equilibrium because they can frequently collide with the nucleons via the neutral weak interaction mediator $Z^{0}$. The average kinetic energy of $\Phi_{u}^{0}$ is therefore $\frac{1}{2}M_{D}v_{D}^{2}=\frac{1}{2}m_{n}v_{n}^{2}=\frac{3}{2}T$. In virtue of $M_{D}\gg m_{n}$, the $\Phi_{u}^{0}$ speed is much slower than the nucleon one.

  As mentioned in the last section, the structure of $\Phi_{u}^{0}$ is that a heavier $\Phi_{R2}^{*}$ confines a lighter $u$ quark via the gluon mediator. This is very similar to the hydrogen atom structure, in which a heavier proton confines a lighter electron via the photon mediator. Therefore, the potential and energy level of $u$ in the inner of $\Phi_{u}^{0}$ can simply be obtained by analogizing ones of electron in the hydrogen atom. In addition, $\Phi_{u}^{0}$ can transition from the ground state to the excited state by means of absorbing collision energy. In accordance with the laws of conservation of momentum and energy, we can write the equations as follows,
\begin{alignat}{1}
&V_{G}=-\frac{4\alpha_{s}}{3\,r}\,,\hspace{0.5cm} E_{n}=-\frac{8m_{u}\alpha_{s}^{2}}{9\,n^{2}}\,,\hspace{0.5cm} \triangle E=E_{2}-E_{1}=\frac{2m_{u}\alpha_{s}^{2}}{3}\,,\nonumber\\
&\mu_{n}(\overrightarrow{v_{r}}- \overrightarrow{v_{r}}')=\overrightarrow{p_{n}}-
 \overrightarrow{p_{n}}'=\overrightarrow{q},\hspace{0.5cm}
 \frac{1}{2}\mu_{n}(v_{r}^{2}-v_{r}'^{2})=\triangle E,\nonumber\\
&q^2=4\mu_{n}E_{r}(1-\frac{\triangle E}{2E_{r}}-\sqrt{1-\frac{\triangle E}{E_{r}}}cos\theta),\hspace{0.5cm} \triangle E=0\:\:\mbox{if}\:\:E_{r}<\triangle E,
\end{alignat}
where $\mu_{n}=\frac{m_{n}M_{D}}{m_{n}+M_{D}}$, $\overrightarrow{v_{r}}=\overrightarrow{v_{n}}-\overrightarrow{v_{D}}$, $E_{r}=\frac{1}{2}\mu_{n}v_{r}^{2}$. $V_{G}$ is one-gluon exchange potential. $\triangle E\approx0.32$ MeV is the lowest transition energy of $\Phi_{u}^{0}$, $q$ is a  momentum transfer of nucleon, and $\theta$ is a scattering angle in the center-of-mass frame. For the collision with $E_{r}<\triangle E$, the collision energy is not enough to excite a transition of $\Phi_{u}^{0}$, this case is an elastic collision, so the kinetic energy loss is vanishing, namely $\triangle E=0$ in (12). For the collision with $E_{r}>\triangle E$, the collision energy is enough to excite a transition of $\Phi_{u}^{0}$, this case is an inelastic collision, so the kinetic energy loss is exactly equal to the transition energy $\triangle E$. However, $\Phi_{u}^{0}$ in the excited state is unstable, it can quickly complete a transition back to the stable ground state by emitting a $\triangle E$-energy $\gamma$ photon. In conclusion, the $\Phi_{u}^{0}$ can actually give out light instead of the dark things in the early stage.

  As the universe temperature decreasing, the relative velocity $v_{r}$ is reducing, accordingly the collisions between $\Phi_{u}^{0}$ and nucleons are becoming rare. Because the collision reaction rate falls faster than the universe expansion rate, the former will be smaller than the latter below a certain temperature. At this point $\Phi_{u}^{0}$ are departure from thermal equilibrium and decoupling, thus the evolution enters the second stage. What follows are a solution to the $\Phi_{u}^{0}$ decoupling temperature.

  At the low energy the collision cross-section of nucleon and $\Phi_{u}^{0}$ is dominated by spin-independent contributions, which arise from the effective vector-vector weak couplings as follows,
\begin{alignat}{1}
&\mathscr{L}_{eff}=-\sum_{q=u,d}[\overline{q}\gamma^{\mu}q][a_{q}\overline{u}\gamma_{\mu}u+b_{q}i(\Phi_{R2}^{\dagger}\partial_{\mu}\Phi_{R2}-\partial_{\mu}\Phi_{R2}^{\dagger}\Phi_{R2})],\nonumber\\
&a_{q}=\frac{g^{2}Q_{q}'Q_{u}'}{M_{Z}^{2}}=\frac{2cos^{2}\theta_{W}Q_{q}'Q_{u}'}{\langle H\rangle^{2}}\,,\hspace{0.3cm}
 b_{q}=\frac{g^{2}Q_{q}'Q_{\Phi_{R2}^{*}}'}{M_{Z}^{2}}=\frac{2cos^{2}\theta_{W}Q_{q}'Q_{\Phi_{R2}^{*}}'}{\langle H\rangle^{2}}\,,\nonumber\\
 &Q'=\frac{1}{cos\theta_{W}}[\frac{1}{2}I^{L}_{3}-Q_{e}sin^{2}\theta_{W}],
\end{alignat}
the above notations are self-explanatory. The collision cross-section is simply calculated by \cite{19}
\begin{alignat}{1}
&\sigma=\frac{\mu_{n}B_{n}^{2}}{512\pi E_{r}}\int_{q^{2}_{min}}^{q^{2}_{max}}F^{2}(q)dq^{2},\nonumber\\
&B_{n}=(A+Z)(a_{u}+b_{u})+(2A-Z)(a_{d}+b_{d}),\hspace{0.3cm} F^{2}(q)=\left(\frac{3j_{1}(qR)}{qR} \right)^{2}e^{-q^{2}s^{2}},
\end{alignat}
where nucleon is denoted by $n(A,Z)$, $F(q)$ is a form factor of $n(A,Z)$, $j_{1}$ is a spherical Bessel function, $R\approx\sqrt{5s^{2}-1.44A}$ fm and $s\approx 1$ fm. $q^2_{min}$ and $q^2_{max}$ are derived from $\theta=0$ and $\theta=\pi$ in (12), respectively. At last the decoupling temperature $T_{D}$ is determined by the formulae as follows,
\begin{alignat}{1}
&\Gamma_{nD}(T_{D})=H(T_{D}),\nonumber\\
&\Gamma_{nD}(T_{D})=\langle\sigma v_{r}\rangle n_{n}=[\frac{\sqrt{2}{n}_{n}}{\sqrt{\mu_{n}}}(\int_{0}^{\triangle E}\sigma_{1}+\int_{\triangle E}^{\frac{\mu_{n}}{2}}\sigma_{2})E_{r}e^{-\frac{E_{r}}{T_{D}}}dE_{r}]/[\int_{0}^{\frac{\mu_{n}}{2}}E_{r}^{\frac{1}{2}}e^{-\frac{E_{r}}{T_{D}}}dE_{r}],\nonumber\\
&n_{n}=g_{n}(\frac{m_{n}T_{D}}{2\pi})^{\frac{3}{2}}e^{-\frac{m_{n}}{T_{D}}},
\end{alignat}
where the heat average is calculated on the basis of Boltzmann distribution of $E_{r}$. $\sigma_{1}$ and $\sigma_{2}$ denotes the cross-section of the elastic collision and one of the inelastic collision, respectively. The integral upper limit $\frac{\mu_{n}}{2}$ is derived from $E_{r}=\frac{1}{2}\mu_{n}v_{r}^{2}$ and $v_{r}<1$. $g_{n}=4$ is the degree of freedom of the nucleon. At this stage the relativistic particles only include electrons, neutrinos and photons, so $g_{*}(T_{D})=12.5$ in (15).

  The $T_{D}$ value will be found by the numerical solution of (15). Undoubtedly, it should be in the region of $1\:\mbox{MeV}\lesssim T_{D}\lesssim 220\:\mbox{MeV}$. After $T<T_{D}$, $\Phi_{u}^{0}$ terminate the collisions with the nucleons due to the departure from thermal equilibrium, namely they are decoupling. Therefore $\Phi_{u}^{0}$ are no longer excited, of course, they can not give out light anymore. From this time on $\Phi_{u}^{0}$ really become the dark matter as so-called name. As the temperature falls to $T\sim 1$ MeV, finally, the universe enters the epoch of BBN \cite{20}.

\vspace{1cm}
 \noindent\textbf{IV. Numerical Results}
\vspace{0.3cm}

  In the section I present the model numerical results. According to the foregoing discussions, the model contains a lot of the new parameters besides the SM ones. In principle the SM parameters have been fixed by the current experimental data, but the non-SM parameters are yet undetermined. The SM parameters involved in the numerical calculations are only the five physical quantities as follows \cite{1},
\begin{alignat}{1}
&\langle H\rangle=174\:\mathrm{GeV},\hspace{0.5cm} m_{u}=2.3\:\mathrm{MeV},\hspace{0.5cm}
 m_{n}=939.6\:\mathrm{MeV},\nonumber\\
&sin^{2}\theta_{W}=0.231,\hspace{0.5cm} \alpha_{s}(1\:\mathrm{GeV})=0.46,
\end{alignat}
where the nucleon mass $m_{n}$ is taken as the neutron one, and the strong gauge coupling coefficient $\alpha_{s}$ is evaluated at the energy scale of $1$ GeV, which is close to the nucleon mass. However, it is enough for our purpose that the non-SM particle masses are only set as some reasonable orders of magnitude rather than fine values, so they are typically chosen as follows,
\begin{alignat}{1}
&M_{N_{1}}=M_{\chi_{1}}=1\times10^{12}\:\mathrm{GeV},\hspace{0.3cm} \frac{M_{N_{2}}}{M_{N_{1}}}=\frac{M_{\chi_{2}}}{M_{\chi_{1}}}=10,\hspace{0.3cm} \frac{M_{N_{3}}}{M_{N_{1}}}=\frac{M_{\chi_{3}}}{M_{\chi_{1}}}=100,\nonumber\\
&M_{L}=1\times10^{8}\:\mathrm{GeV},\hspace{0.3cm} M_{1}=1\times10^{3}\:\mathrm{GeV},\hspace{0.3cm}
 M_{2}\approx M_{D}=800\:\mathrm{GeV}.
\end{alignat}
Here I only suppose that $M_{N_{1}}$ and $M_{\chi_{1}}$ are smaller than $T_{reheat}$ for simplicity, so only the decays of $N_{1}$ and $\chi_{1}$ in (7) and (8) need be considered. The Yukawa couplings, $Y_{N},Y_{L},Y_{1},Y_{2}$, actually contain a great deal of the flavour parameters. They have larger freedoms since the flavour structures are as yet unknown. In view of (7) and (8), the Yukawa matrix elements can simply be taken as follows,
\begin{alignat}{1}
&(Y_{N}^{\dagger}Y_{N})_{11}=5\times10^{-2},\hspace{0.3cm}
 (Y_{L}^{\dagger}Y_{L})_{11}=(Y_{1}^{\dagger}Y_{1})_{11}=(Y_{2}^{\dagger}Y_{2})_{11}=1\times10^{-6},\nonumber\\
&\mbox{Im}(Y_{L}^{\dagger}Y_{L})_{1j}^{2}=4\times10^{-8},\hspace{0.3cm}
 \mbox{Im}(Y_{1}^{\dagger}Y_{1})_{1j}^{2}=\mbox{Im}(Y_{2}^{\dagger}Y_{2})_{1j}^{2}=1\times10^{-14},
\end{alignat}
where $j=1,2,3$. $(Y_{N}^{\dagger}Y_{N})_{11}$ and $\mbox{Im}(Y_{L}^{\dagger}Y_{L})_{1j}^{2}$ are relative sensitive to fitting $m_{\nu}$ and $\eta_{B}$, so their fine values are given, but the other parameters are only fixed to the suitable orders of magnitude. In addition, $\mbox{Im}(Y_{N}^{\dagger}Y_{N})_{1j}^{2}$ is absent because $N_{1}\rightarrow l+\widetilde{H}^{*}$ is actually not out-of-equilibrium.

  The above values of the parameters are based on an overall consideration, namely, they not only satisfy the model consistency and the experimental limits, but also are typical in the parameter space. Firstly, (17) and (18) are put into (8) and (9), we can obtain the results as follows,
\begin{alignat}{1}
&\frac{\Gamma_{1}}{H(M_{N_{1}})}\approx639,\hspace{0.3cm} \frac{\Gamma_{2}}{H(M_{N_{1}})}\approx 0.038,\hspace{0.3cm} \frac{\Gamma_{3}}{H(M_{\chi_{1}})}=\frac{\Gamma_{4}}{H(M_{\chi_{1}})}\approx 0.019,\nonumber\\
&T_{LF}\approx 1222\:\mathrm{GeV}.
\end{alignat}
These clearly show that relative to the universe expansion, $N_{1}\rightarrow l+\widetilde{H}^{*}$ is very fast and the other three decays of $N_{1}$ and $\chi_{1}$ are very weak, in other words, the other three decays are serious out-of-equilibrium but the first decay is full in thermal equilibrium. Indeed, $\Gamma_{1}<H(M_{N_{1}})$ is very difficult to be satisfied in view of the constraint of the neutrino mass (5). Therefore, we can reasonably infer that $\kappa_{1}\approx 0$ and $\kappa_{2,3,4}\approx 1$ in (11). We can thus draw a conclusion that the lepton asymmetry essentially arises from $N_{1}\rightarrow q+\Phi_{L}^{*}$ rather than $N_{1}\rightarrow l+\widetilde{H}^{*}$, while the DM asymmetry entirely stems from $\chi_{L}\rightarrow d_{R}+\Phi_{R1}^{*}$ and $\chi_{L}\rightarrow u_{R}+\Phi_{R2}^{*}$. On the other hand, $T_{LF}\approx 1222$ GeV exactly meets our expectation, which can guarantee that the sphaleron processes are put in effect smoothly.

  Secondly, by the calculation of (7) and (11), we can obtain the asymmetries of DM and baryon and the ratio of their abundance at the present-day, namely
\ba
 \eta_{D}\approx 3.6\times10^{-12},\hspace{0.5cm} \eta_{B}\approx 6.1\times10^{-10},\hspace{0.5cm}
 \frac{\Omega_{D}}{\Omega_{B}}\approx 5\,.
\ea
$\eta_{D}$ which is as yet undetected is mainly subject to $Y_{1}^{\dagger}Y_{1}$ and $Y_{2}^{\dagger}Y_{2}$, however, it is believed to be two orders of magnitude smaller than $\eta_{B}$. Since $\eta_{D}$ and $M_{D}$ are together in charge of $\frac{\Omega_{D}}{\Omega_{B}}$, $M_{D}$ can vary in a certain area. In short, these results are very well accordance with the current data of the universe observations \cite{21}.

  Lastly, put (16) into (12)--(15) and fulfil the numerical calculations, we can obtain $\triangle E$ and $T_{D}$. In addition, no $\gamma$ photons are emitted in the experiments of the direct detection of the DM so far, therefore we can infer that the average kinetic energy of the DM in the present-day universe should be less than $\triangle E$, namely $\frac{1}{2}M_{D}v_{D}^{2}<\triangle E$ in which $v_{D}\approx220\:km/s\approx7.33\times10^{-4}\:c$ is the average speed of the DM at the present-day \cite{22}. Thus the mass upper limit is derived as $M_{D}<2\triangle E/v_{D}^{2}$. All the important results are summed up as follows,
\ba
\triangle E\approx 0.32\:\mathrm{MeV},\hspace{0.5cm} T_{D}\approx 130\:\mathrm{MeV},\hspace{0.5cm} M_{D}<1207\:\mathrm{GeV}.
\ea
It should be stressed that the above results only depend on the five parameters in (16), and they are approximately independent of $M_{D}$ due to $\mu_{n}\approx m_{n}$. The lower limit of $M_{D}$ should be provided by the collider searches. In a word, (21) are three interesting and important predictions of the model.

  Here I give a brief discussion about the model test. Firstly, $\triangle E\approx 0.32$ MeV is in the $\gamma$-energy range, which is far higher than the visible region and beyond the X-band. I suggest two ways to detect the $\gamma$ photons of $0.32$ MeV. One is that the $\gamma$ photons can arise from the collisions of some galaxies and the DM by chance, so we can search them in the cosmic observations. The other one is that we can use the neutron beam with $1$ MeV kinetic energy to collide the DM, then we can detect the $\gamma$ photons through the neutron scattering.

  Secondly, $T_{D}\approx 130$ MeV exactly fits what we expected, this is purely a prediction of the model but by no means a coincidence. In fact, it is the DM decoupling temperature from the collisions with neutrons. The DM decoupling from the collisions with protons is actually earlier than one from the collisions with neutrons because the collision cross-section with proton is a few smaller than one with neutron. Evidently, the period from the QCD phase transition to the BBN is indeed divided into two phases by $T_{D}\approx 130$ MeV. In the phase of $T_{D}<T<T_{QCD}$, the DM can emit the $\gamma$ photons of $0.32$ MeV through the collisions with nucleons. In the phase of $T<T_{D}$, the DM are decoupling and stop to emit the $\gamma$ photons completely, so they become the dark matter.

  Lastly, we can search the color scalar boson $\Phi_{R2}$ at the colliders on account of $M_{D}<1207$ GeV. On the basis of the model interactions, a pair of $\Phi_{R2}$ can be produced by three ways as follows,
\begin{alignat}{1}
 &e^{-}+e^{+}\rightarrow\gamma\rightarrow \Phi_{R2}+\Phi_{R2}^{*},\hspace{0.5cm}
  p+\overline{p}\rightarrow G\rightarrow \Phi_{R2}+\Phi_{R2}^{*},\nonumber\\
 &p+p\rightarrow G+G\rightarrow \Phi_{R2}+\Phi_{R2}^{*}.
\end{alignat}
However, they eventually annihilate into photons instead of any decay products. The first process can be accomplished at the future lepton-antilepton collider as the ILC \cite{23}. This is also the best efficient method to measure $\Phi_{R2}$. The last process can be searched at the present LHC \cite{24}. We are looking forward to the relevant results. Although all of the suggested searches are some large challenges, the model is feasible and promising to be tested in near future.

\vspace{1cm}
 \noindent\textbf{V. Conclusions}
\vspace{0.3cm}

  In the paper, I suggest a simple model of the asymmetric dark matter and leptogenesis. The model is based on the SM gauge groups and the $Z_{2}$ discrete symmetry. The baryon number conservation is incidental but the lepton number is not conserved. The new particles in the model are the two super-heavy Majorana fermions and the three color scalar bosons. In virtue of the $CP$ violation and out-of-equilibrium, the lepton asymmetry is essentially generated by the decay chain of $N_{R}\rightarrow q+\Phi_{L}^{*}$ and $\Phi_{L}^{*}\rightarrow \overline{q}+\overline{l}+\widetilde{H}$, which is then converted into the baryon asymmetry by the sphaleron processes. The two decay chains of $\chi_{L}$ in (6) eventually lead to the asymmetries of $\Phi_{R2}^{*}$ and $u$, which are then combined into the asymmetric stable $\Phi_{u}^{0}$ after the QCD phase transition. At the early stage $\Phi_{u}^{0}$ are in thermal equilibrium through the collisions with the nucleons, moreover, they can emit the $\gamma$ photons with $0.32$ MeV energy. At $T_{D}\approx 130$ MeV, $\Phi_{u}^{0}$ are decoupling completely, from then on they no longer give out light and become the dark matter. The model not only accounts for the origins of the ADM and matter-antimatter asymmetry, and the close relationship of the both, but also it elaborates the dark matter nature. In particular, the model gives the three important predictions of the ADM, namely the transition energy of $0.32$ MeV, the decoupling temperature of $130$ MeV, and the mass upper limit of $1207$ GeV. Finally, these ideas and predictions can certainly provide some guides for the future experimental search, the model is expected to be tested in near future.

\vspace{1cm}
 \noindent\textbf{Acknowledgments}
\vspace{0.3cm}

  I would like to thank my wife for her large helps. This research is supported by chinese universities scientific fund under Grant No. WK2030040003.

\end{document}